\documentclass[conference]{IEEEtran}
\usepackage{fancyhdr} 
\IEEEoverridecommandlockouts

\usepackage{cite}
\usepackage{amsmath,amssymb,amsfonts}
\usepackage{algorithmic}
\usepackage{graphicx}
\usepackage{textcomp}
\usepackage{multirow}
\usepackage{xcolor}
\usepackage[hyphens]{url}
\usepackage[caption=false,font=footnotesize]{subfig}
\usepackage{hyperref}
\usepackage{cleveref}
\usepackage{booktabs}
\usepackage{tabularx}

\begin{document}
% \IEEEoverridecommandlockouts
\IEEEpubid{\makebox[\columnwidth]{979-8-3195-1221-5/26/\$31.00 \copyright2026 IEEE \hfill} \hspace{\columnsep}\makebox[\columnwidth]{}}

\title{To SIMD or Not to SIMD: Comparing RVV vs. CGRA SoCs for Non-Linear Functions in AI}
\author{\IEEEauthorblockN{Sameer Deshmukh}
\IEEEauthorblockA{\textit{Fujitsu Research of Japan} \\
\textit{Fujitsu Limited}\\
Kawasaki, Japan \\
deshmukh.sameer@fujitsu.com}
\and
\IEEEauthorblockN{Naoya Mizuki}
\IEEEauthorblockA{\textit{Fujitsu Research of Japan} \\
\textit{Fujitsu Limited}\\
Kawasaki, Japan \\
mizuki.naoya@fujitsu.com}
\and
\IEEEauthorblockN{Takumi Honda}
\IEEEauthorblockA{\textit{Fujitsu Research of Japan} \\
\textit{Fujitsu Limited}\\
Kawasaki, Japan \\
honda.takumi@fujitsu.com}
}
%%%%%%%%%%%%%%%%%%%%%%%%%%%%%%%%%%%%%%%%

\maketitle

\thispagestyle{fancy}
\lhead{}
\rhead{}
\chead{}
\lfoot{\footnotesize{
SC26 Workshops, November 15-20, 2026, Chicago, Illinois, USA
\newline 979-8-3195-1221-5/26/\$31.00 \copyright 2026 IEEE}}
\rfoot{}
\cfoot{}
\renewcommand{\headrulewidth}{0pt}
\renewcommand{\footrulewidth}{0pt}

\begin{abstract}
Non-linear functions are an important class of operations in modern large language
models. However, chip vendors building for AI have so far relied
on traditional vector or scalar processing units for non-linear functions.
The performance of these units has not kept pace with matrix multiplication
units (MMUs), which can be found in both high-performance and edge-computing devices
from almost every vendor. Reconfigurable units such as FPGAs and CGRAs have
been proposed in the literature, but have not been systematically compared with
traditional units backed by MMUs. In this paper, we build two SoCs: one using
an RVV vector processor and the other using a latency-insensitive (\emph{elastic}) CGRA, both coupled with
an MMU and driven by a single-core CPU. We then propose mappings for common feed-forward networks
on our SoCs and find that the CGRA reaches a maximum speedup in clock cycles of \textbf{7.12x} over RVV on the
activation stage of simple point-wise non-linear functions like $\mathrm{SiLU}$, whereas the RVV
reaches a maximum speedup of \textbf{9.8x} over the CGRA on complex non-linear functions such as
$\mathrm{softmax}$. Over the enclosing feed-forward block the corresponding maxima are
\textbf{3.16x} and \textbf{4.24x}.
\end{abstract}

\section{Introduction}
\label{sec:introduction}

The death of Moore's law has opened up new frontiers for computer architects to build
designs for improved power and efficiency to serve the growing demands of artificial 
intelligence (AI) and scientific computing. AI models such as LLMs and CNNs
are dominated by the dense general matrix multiplication (GEMM), which has led
an increasing number of hardware vendors to include dedicated matrix multiplication 
units on their hardware\cite{nvidia2017v100,amd2020cdna,prabhakar2024sambanova,intel2024amx,arm2022sme}.

While GEMM is the most dominant routine in common AI models\cite{kaplan2020scaling,sze2017efficient,jouppi2017tpu}, 
non-linear functions such as $\mathrm{SiLU}$\cite{Elfwing2017silu} and $\mathrm{softmax}$\cite{Stevens2022softermax} play a pivotal
role in their accuracy. These non-linear functions and others like 
them\cite{zhang2019rmsnorm,ba2016layernorm,dauphin2017glu,hendrycks2016gelu,nair2010rectified} 
require the application of functions such as exponential, logarithm,
division, inequalities, extrema and trigonometric functions, which the design of a
typical matrix multiplication unit\cite{genc2021gemmini,jouppi2017tpu,kung1978systolic} is not able to support.
Therefore, a non-linear
function must be computed by a separate compute unit such as a vector or scalar processing
unit. Examples of such complementary designs include SFU units\cite{oberman2005interpolator} 
on Nvidia GPUs and SIMD vector processing units on Intel Granite 
Rapids\cite{anderson2026correctlyrounded,intel2025graniterapidsd}.

GEMM performance has seen a significant increase over the past decade compared to
non-GEMM performance\cite{matsuoka2023myths}, especially for low precision data types such as
fp8\cite{micikevicius2022fp8} and bfloat16\cite{kalamkar2019bfloat16}. The computation of the 
non-linearity, on the other hand, is still typically
performed in single-precision 32-bit floating-point data types\cite{micikevicius2022fp8}.
The gap between matrix multiplication units using low precision and other arithmetic units
on modern GPUs has become so significant that FlashAttention 4\cite{zadouri2026flashattention4} 
simulates the exponential in software because of the
performance difference between the tensor cores and the SMs. Clearly, there is
a need for a hardware unit for non-linear function computation that is able to stay in step with advances in
matrix multiplication performance.

Reconfigurable hardware units such as 
FPGAs\cite{lu2023autolut,hong2022dfx, li2020ftrans, huang2024edgellm, jeong2023fpga, li2023reconfigurable} and 
CGRAs\cite{Qin2025picachu,sankaralingam2022mozart, lee2021npcgra} have 
been proposed as alternative ways to compute non-linear functions by taking advantage of their
superior pipelining. PICACHU\cite{Qin2025picachu} combines a CGRA with a 
matrix multiplication unit to test deep learning workloads, showing improved performance compared
to just a CGRA. However, a proper comparison showing the pros
and cons of a CGRA coupled with a matrix multiplication unit vs. a traditional vector or scalar processing
unit combined with a matrix multiplication unit is lacking in the literature. A fair comparison against existing computer
architectures can show the specific strengths and weaknesses of a CGRA companion in terms of
achieved performance, programmability and mapping effort. Comparing
against an existing architecture from a major vendor does not allow for a fair comparison due to barriers
posed by vendor-specific proprietary IP and the highly complex nature of commercially available hardware.
Moreover, one cannot easily compare implementations of equivalent functionality in the CGRA and in a traditional
vector or scalar processing unit to deeply understand the advantages or disadvantages of either.

This paper aims to fill this gap by proposing two SoCs: Matrix Engine-Vector Engine (ME-VE) and 
Matrix Engine-Reconfigurable Engine (ME-RE). The ME-VE combines a Saturn\cite{Zhao2024saturn} RISC-V 
vector processor\cite{riscvvector10} with a Gemmini matrix multiplication\cite{genc2021gemmini} unit,
whereas the ME-RE combines an elastic RIKEN CGRA\cite{Adhi2022exploration} with the same Gemmini matrix
multiplication unit. Both SoCs are driven by a single-core, in-order RISC-V Rocket CPU\cite{asanovic2016rocket}
and feature exactly the same memory hierarchy. Both SoCs make use of fully open-source hardware stacks --- CGRA-ME\cite{Ragheb2024cgrame2} for building interfaces to the CGRA and Chipyard\cite{amid2020chipyard}
for building all other SoC components such as memory controllers, CPU and buses.

In this paper, we compare commonly used feed-forward network (FFN) blocks from the latest model
architectures\cite{Grattafiori2024llama3,qwen2024qwen2,kimi_team2026kimik3} that combine a linear
function (matrix multiplication) with a non-linear function on our proposed SoCs. Specifically, we
make the following contributions:
\begin{enumerate}
    \item Build two SoCs using identical memory hierarchies and CPU drivers, differing only in
    their use of a vector processing unit or a reconfigurable unit for a fair comparison between
    a vector processor and a CGRA.
    \item Show that on the non-linear stage in isolation the CGRA reaches a maximum speedup of
    \textbf{7.12x} over the vector processor for $\mathrm{SiLU}()$, whereas the vector processor
    reaches a maximum speedup of \textbf{9.8x} over the CGRA for $\mathrm{softmax}$. Amortised over
    the enclosing FFN block, which also contains the matrix multiplications, these maxima become
    \textbf{3.16x} and \textbf{4.24x} respectively.
    \item Provide detailed benchmarks of the SoCs to study the trade-offs between a vector processor and a CGRA.
\end{enumerate}

\section{Background}
\label{sec:background}

Linear functions such as matrix multiplication are frequently combined with a non-linear function such as
$\mathrm{SiLU}$\cite{Elfwing2017silu} or $\mathrm{softmax}$\cite{Blanchard2021logsumexp} in the latest 
AI models\cite{kimi_team2026kimik3,Grattafiori2024llama3}. Our goal is to find the performance differences
between a vector processor and a reconfigurable array for combinations of such linear and non-linear 
functions.
\Cref{tab:functions-tested} shows the functions we benchmark in this paper. The SwiGLU contains 3 matrix multiplications and a single non-linear
function $\mathrm{SiLU}$. It is characterized by element-wise looping over the input values where every value is accessed
exactly $O(1)$ times. Attention, on the other hand, consists of 5 matrix multiplications and a $\mathrm{softmax}()$ function
call. The $\mathrm{softmax}$ function consists of a reduction loop, followed by a scalar reciprocal and finally an
element-wise division. Thus, these two non-linear functions have very different computational 
characteristics, showing distinct memory access patterns and amounts of computation.

\begin{table}
\centering
\caption{Combinations of linear and non-linear operators that we test.}
\label{tab:functions-tested}
\begin{tabular}{c|c}
\textbf{Name} & \textbf{Equation} \\
\hline
SwiGLU\cite{Shazeer2020glu,Grattafiori2024llama3, qwen2024qwen2}  & $(\mathrm{SiLU}(XW_{gate}) \odot XW_{up})W_{down}$ \\
Attention\cite{kimi_team2026kimik3,Grattafiori2024llama3,Vaswani2017attention} & $\mathrm{softmax}\!\left(\frac{(XW_Q)(XW_K)^{\top}}{\sqrt{N}}\right)XW_V$ \\
\end{tabular}

\end{table}

The matrix multiplications with the $W_{gate}$, $W_{up}$ and $W_{down}$ weight matrices are performed on Gemmini. 
The non-linear functions such as $\mathrm{SiLU}()$ and $\mathrm{softmax}()$ are computed on the other accelerator,
which can either be a vector processor or a CGRA depending on the SoC being tested.

Our testbed consists of two SoCs: ME-VE and ME-RE, using the vector processing unit and the reconfigurable
array, respectively. Both SoCs are driven by an in-order, single-core RISC-V Rocket CPU\cite{asanovic2016rocket}. 
They also use exactly the same memory hierarchy, consisting of a 256 MB DRAM (simulated using 
DRAMSim2\cite{Rosenfeld2011dramsim2}), a 512 kB, 8-way SiFive inclusive L2 cache\cite{sifivel2cache,amid2020chipyard} and an
L1 cache of size 32 kB within the Rocket CPU core. The SoC components connect with their respective buses using TileLink\cite{SiFive2017DiplomaticDP}, and the interface between the memory bus and the DRAM uses the AXI4 protocol\cite{arm_axi4_spec}. Specific details of the ME-VE and ME-RE are elaborated in \Cref{sec:matmul-with-vector-engine} and \Cref{sec:matmul-with-cgra}, respectively.

\section{Matrix engine with vector processor (ME-VE)}
\label{sec:matmul-with-vector-engine}

\begin{figure}
    \centering
    \includegraphics[width=\linewidth]{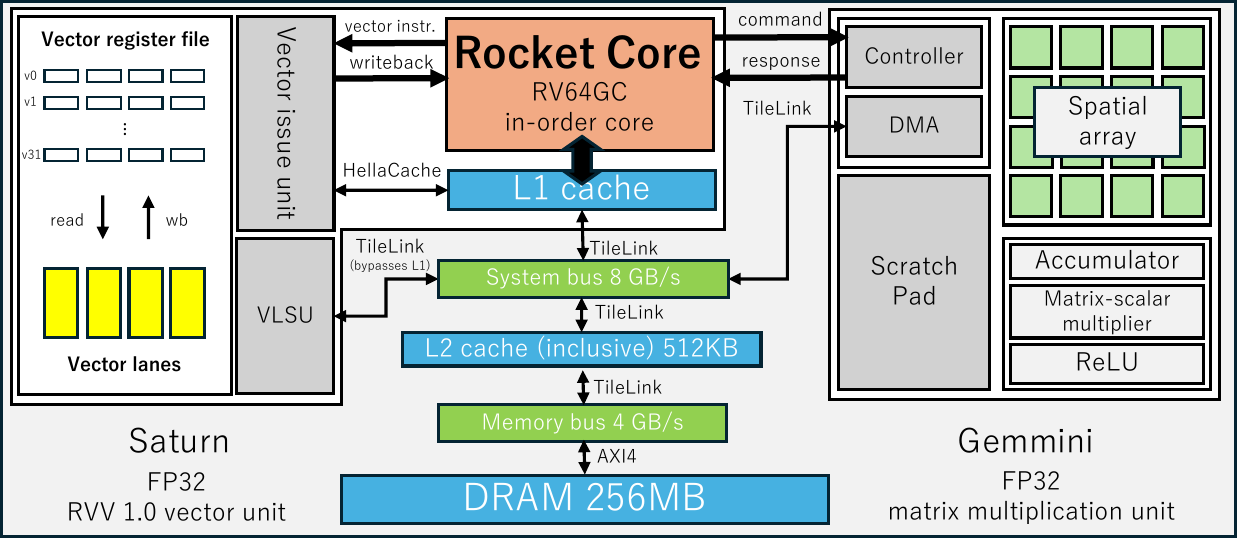}
    \caption{The ``ME-VE'' SoC. We combine a Saturn RISC-V vector processor with Gemmini using a Rocket CPU to simulate the combination of a matrix multiplication accelerator with a vector processor.}
    \label{fig:rocket-core-with-rvv}
\end{figure}

\Cref{fig:rocket-core-with-rvv} shows the structure of the ME-VE SoC.
The vector unit is the Saturn vector unit\cite{Zhao2024saturn} using the RVV 1.0 instruction set\cite{riscvvector10}. 
It is coupled with a Gemmini\cite{genc2021gemmini} matrix multiplication processor.
Unlike Gemmini, Saturn is not a memory-mapped or RoCC-attached accelerator; it is bound
to the CPU core, so the vector instructions are fetched, decoded and committed by the scalar core's own pipeline. Saturn co-exists with the CPU on the same tile.

Saturn is configured to be a 128-bit wide vector processor. It contains 32 128-bit
registers to store a total of 512 bytes. It is fed by a 128-bit wide datapath from
the registers into the compute units. It has two independent datapaths into memory: the first is a dedicated TileLink\cite{SiFive2017DiplomaticDP} interface that bypasses the L1 cache and attaches directly to the system bus, and the second is a connection to the L1 cache through
a HellaCache\cite{asanovic2016rocket} port. Although the Rocket CPU is an in-order CPU, the vector instructions committed by the CPU to Saturn are executed out of order using a separate out-of-order execution pipeline.
It features a 4-stage pipeline.

The RVV specification allows for flexibility in configuration. We configure the
datapath between the registers and the compute ports to be 128 bits to make
full utilization of the SIMD units. We tried different configurations of the
register grouping parameter \texttt{LMUL}\cite{Shi2026closergap}
and set it to $4$, so that a single vector instruction operates on a group of four architectural
registers rather than one. With a 128-bit \texttt{VLEN} this makes one \texttt{vle32.v} move
16 FP32 elements; the datapath itself remains 128 bits wide, so the access is spread over four
cycles rather than widened. A large \texttt{LMUL}
shows significant performance benefits for a small problem size but no
noticeable change as the problem size becomes larger.

This setting is experimentally
found to be the fastest because it can saturate the load pipeline of the vector unit
and significantly reduce the number of instructions, thus reducing the fixed cost per
instruction.

\section{Matrix engine with CGRA (ME-RE)}
\label{sec:matmul-with-cgra}

\begin{figure}
    \centering
    \includegraphics[width=\linewidth]{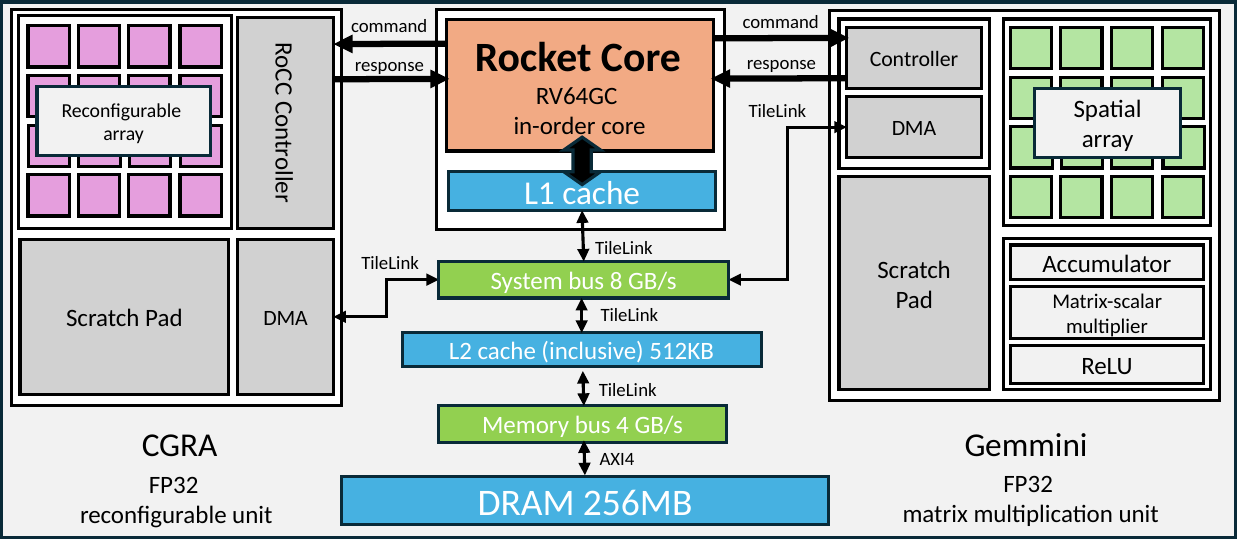}
    \caption{The ``ME-RE'' SoC. We combine a CGRA-ME elastic RIKEN architecture with Gemmini using a Rocket CPU to simulate the combination of a matrix multiplication accelerator with a reconfigurable array.}
    \label{fig:cgra-me-with-gemmini}
\end{figure}

\Cref{fig:cgra-me-with-gemmini} shows our SoC architecture combining a matrix multiplication
unit (Gemmini\cite{genc2021gemmini}) with a reconfigurable array (CGRA).

\subsection{Gemmini specifications}

Gemmini is configured as a 4x4 spatial array for FP32 computation.
It is directly connected to an inclusive L2 cache which is shared with the Rocket core CPU and 
the CGRA. The DMA engine in Gemmini\cite{genc2021gemmini} allows it to directly access memory via the L2 cache.
The Rocket CPU can control Gemmini using the RoCC protocol\cite{asanovic2016rocket}. 
The Rocket CPU and Gemmini connections use the default configuration
provided by Chipyard\cite{amid2020chipyard}.

\subsection{CGRA specification}

The CGRA accelerator is generated with CGRA-ME\cite{Ragheb2024cgrame2} using the RIKEN elastic architecture\cite{Adhi2022exploration}
as the underlying CGRA architecture. An algorithm is first mapped onto
the CGRA fabric using an appropriate mapping algorithm, and the generated mapped CGRA fabric is 
connected to Chipyard via the RoCC controller. The Rocket CPU\cite{asanovic2016rocket} sends
commands to the RoCC controller\cite{amid2020chipyard},
which controls the scratchpad and reconfigurable array. The scratchpad and reconfigurable array are internal to the CGRA unit and do not have their
own bus. A DMA unit (controlled by the RoCC controller) issues load / store commands
to the inclusive L2 cache over a TileLink bus\cite{SiFive2017DiplomaticDP}.
The RIKEN CGRA architecture used throughout this study consists of a 4x4 grid of
PEs using pipelined FloPoCo\cite{Dedinechin2011flopoco} FP32 cores. Each PE features a 
multiplication (FMUL), addition (FADD),
division (FDIV), exponential (FEXP) and square root (FSQRT) unit.

Per-operator latencies are read from the generated cores rather than assumed:
each operator's pipeline depth was obtained by counting its registered stages in
the Verilog produced by FloPoCo~\cite{Dedinechin2011flopoco} (stage registers are named
\texttt{\_dN}, so the maximum $N$ is the depth), giving 3 cycles for FADD and
FEXP, 6 for FMUL, 10 for FSQRT and 15 for FDIV. These agree exactly with the
latency table CGRA-ME~\cite{Ragheb2024cgrame2,chin2017cgrame} uses to schedule the dataflow graph,
and are independently corroborated by the per-opcode taps in the functional
unit's valid-shadow shift register. FloPoCo pipelines an operator to meet a
target frequency rather than to a specified depth, inserting registers wherever
the accumulated combinational delay would exceed the per-stage
budget~\cite{Isotan2017automating}; a higher target therefore yields a deeper pipeline for
identical logic, and FEXP is shallow because it is evaluated as a table-driven
polynomial rather than iteratively~\cite{Detrey2007exp}. The cores are
pre-generated FloPoCo artifacts vendored from prior work, and their generation
banners record four different device-and-frequency targets---Kintex7 at
200\,MHz for FADD, StratixV at 500\,MHz for FMUL, StratixV at 350\,MHz for FDIV,
and Virtex6 at 400\,MHz for FSQRT---so pipeline depth reflects the per-stage
timing budget each core was generated against rather than relative operator
complexity or a uniform budget for this fabric. All four are FPGA delay models,
whose characteristics differ substantially from the ASIC technologies this SoC
targets~\cite{Kuon2007measuring}.

This affects our results principally through depth, and that effect is bounded.
The two softmax paths of \Cref{sec:softmax-mapping} share the prefix
FMUL$\to$FADD$\to$FEXP, which accounts for $6+3+3=12$ cycles; the reduction path
adds a terminal FADD for 15 cycles in total and the element-wise path a terminal
FMUL for 18. Of these operators only FADD and FEXP were generated below
500\,MHz, so the sub-500\,MHz depth is 9 of 15 cycles on the reduction path and
6 of 18 on the element-wise path. Regenerating them against a 2\,ns rather than
a 5\,ns per-stage budget can at most scale that depth by the budget ratio of
$2.5$, adding 13.5 and 9 cycles to the two paths respectively, or 11.25 cycles
averaged over the two passes an element makes. Because the fabric sustains an
initiation interval of one~\cite{Rau1994iterative}, additional depth is paid once per
fabric invocation as pipeline fill rather than per token---the standard
latency-versus-period trade~\cite{Leiserson1991retiming}. One invocation is one chunk
streamed once, so a row of $N$ elements costs $2\lceil N/24 \rceil$ invocations and
the whole problem $2N\lceil N/24 \rceil$: 32, 128, 384 and 1536 invocations at
$N=16$, 32, 64 and 128. The total cost is therefore 11.25 cycles times the
invocation count, or under 1.5\% of the softmax stage at every problem size
measured (1.45\%, 1.44\%, 1.21\% and 1.18\%).
A second channel is \emph{not} covered by that argument, and we state it here as a limitation of
this study. The generation targets fix the \emph{ratio} of depths between operators, not only their
absolute values, and it is the ratio that makes the two branches of the $\mathrm{SiLU}$ DFG 27 and
6 cycles deep. That imbalance is what sets the elastic buffer depth, and it is also the origin of
the operand-duplication workaround of \Cref{sec:performance-analysis-attention}, which costs a
factor of two on the softmax fabric rather than a per-invocation fill. Because the effect is
multiplicative rather than additive, the pipeline-fill bound above does not contain it: a fabric
whose operators were all generated against one uniform target could exhibit a different branch
imbalance and therefore a different duplication cost. Quantifying this would require regenerating
all five operators against a single target and repeating the sweep, which we have not done.
Both SoCs are configured at the same clock and we report cycle counts rather
than wall-clock time, so no other channel exists by which the generation targets
could affect the comparison. Synthesis and static timing analysis, which would
establish the achieved frequency of each SoC, are left as future work.

Because the fabric is elastic, a longer path in a mapped DFG exerts back
pressure on any shorter path it joins\cite{Lana2021placement,Lines2006slack}, so an elastic buffer
of an appropriate depth must be placed on the shorter path to drain it. We use an elastic buffer
depth of $32$ throughout this study, for both the $\mathrm{SiLU}$ and the $\mathrm{softmax}$
mappings. The mapping is done using a simulated
annealing mapper that comes built-in with CGRA-ME\cite{chin2017cgrame}.

\subsubsection{Data streaming}

The CGRA uses a DMA unit to directly stream data to and from the DRAM using the inclusive L2
cache. The DMA unit loads the data from the L2 into the scratchpad for use by the reconfigurable
array. It uses a scratchpad as a staging area for the data before and after processing.  
Three buffers in the scratchpad are used for storing two operands and a result, each of size 512 bytes.
The total scratchpad size is 1536 bytes. The CGRA computation adopts a chunking approach
for streaming data from the memory. Although each buffer of 512 bytes is capable of storing up to
128 FP32 numbers, it is split into two halves of 64 FP32 numbers, so that one half can in principle
be refilled while the other is being streamed. Within a half, each element occupies two slots in
order to account for the fork-skew\cite{Carloni2001} in the SiLU DFG, which fixes the chunk at
$64/2 = 32$ operand positions. Eight of those positions are consumed by the run prologue described
in \Cref{sec:performance-analysis-attention}, so a chunk carries \textbf{32 operand positions of
which 24 are payload elements}; ``chunk of 24'' throughout \Cref{sec:results} refers to this
payload count. The Gemmini scratchpad has a size of 256 kB and the accumulator within
Gemmini has a capacity of 64 kB. 

% Specify tilelink interfaces such as width. scratchpad size. what determines the scratchpad

\subsubsection{Controlling the CGRA from the CPU}

The RISC-V specification reserves 4 major opcodes specifically for user-defined,
application-specific hardware accelerators\cite{riscv-unpriv-isa} (custom0, custom1, custom2 and
custom3). Gemmini\cite{genc2021gemmini,amid2020chipyard} is controlled using the custom3 opcode
(\texttt{0x7b}) and the CGRA using the custom2 opcode (\texttt{0x5b}).

The RISC-V R-type assembly instruction used for this custom2 opcode has the
form ``\texttt{.insn r 0x5b, funct3, funct7, rd, rs1, rs2}''. Here, \texttt{0x5b}
denotes the custom2 instruction\cite{riscv-unpriv-isa}, \texttt{funct3}
is a 3-bit sub-field that selects between different variants of the major opcode, and
\texttt{funct7} is a 7-bit field which specifies
the precise action that this instruction must perform. \texttt{rd}, \texttt{rs1} and
\texttt{rs2} denote the destination register, operand1 and operand2, respectively.
We use a simple instruction set of 8 instructions to control the CGRA from the CPU,
encoded in the \texttt{funct7} values. \Cref{tab:swiglu-custom2-instructions} shows the 
value carried by \texttt{funct7} and a description of the corresponding instruction for 
the SiLU mapping.

\begin{table}[t]
  \centering
  \caption{R-type \texttt{custom2} instruction encoding used to drive the CGRA
           accelerator.}
  \label{tab:swiglu-custom2-instructions}
  \begin{tabularx}{\linewidth}{@{}clX@{}}
    \toprule
    \textbf{funct7} & \textbf{Name} & \textbf{Purpose} \\
    \midrule
    0 & \texttt{CFG} & Scan the configuration bitstream into the fabric --- run once before any compute. \\
    1 & \texttt{SET} & Write one configuration register (scratchpad pointer, collector latency, or a DMA base address). \\
    2 & \texttt{FILL} & Start a non-blocking DMA read that fills one scratchpad half with operands. \\
    3 & \texttt{PUSH2} & Fill the scratchpad from core registers, two operand pairs per instruction. \\
    4 & \texttt{RUN} & Arm the streaming engine over one scratchpad half and return immediately. \\
    5 & \texttt{WAIT} & Ask whether the engine or the DMA has finished, either blocking or polling. \\
    6 & \texttt{STAT} & Read one counter, one trace entry, or one staged result via a single readout port. \\
    7 & \texttt{PUSH1} & Fill the scratchpad one operand pair per instruction. \\
    \bottomrule
  \end{tabularx}
\end{table}

\subsubsection{Mapping the \texorpdfstring{$\mathrm{SiLU}$}{SiLU}} 
\label{sec:silu_mapping}

\begin{figure}[!t]
    \centering
    \subfloat[$\mathrm{SiLU}$\label{fig:silu-cgra-mapping}]{%
        \includegraphics[width=0.368\textwidth]{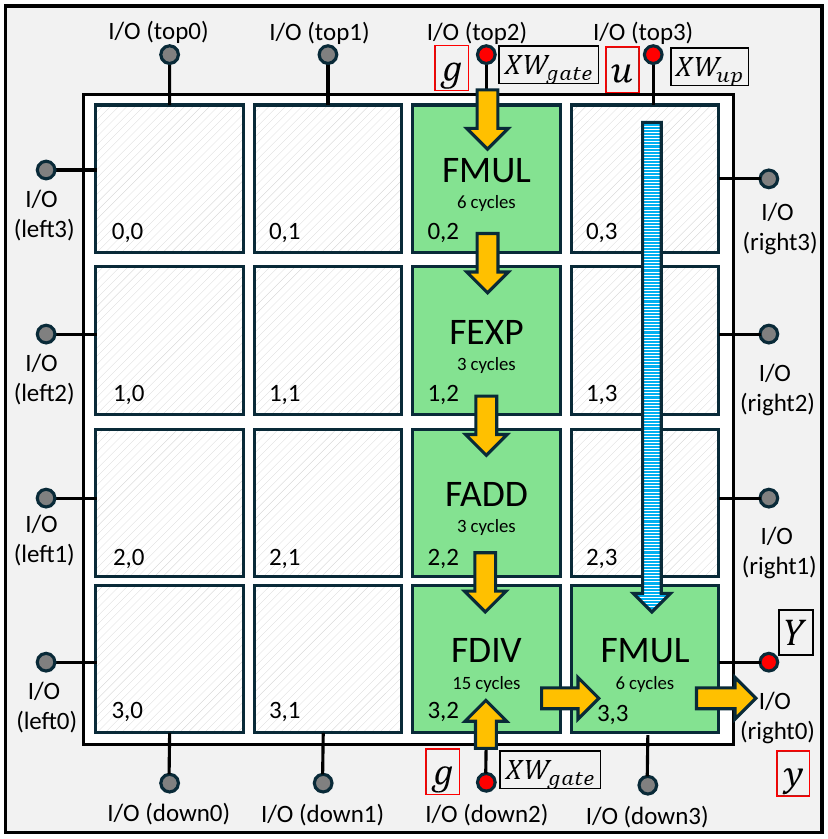}}
    \\
    \subfloat[Softmax\label{fig:softmax-mapping}]{%
        \includegraphics[width=0.368\textwidth]{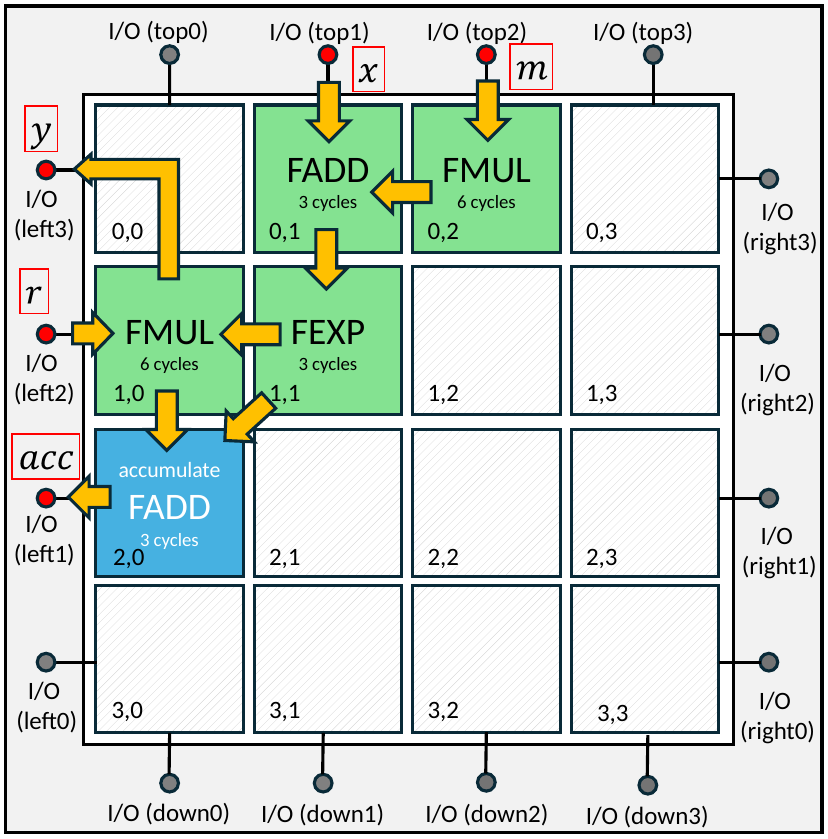}}
    \caption{Mapping of non-linear operators onto our 4x4 elastic RIKEN CGRA:
    (a)~the $\mathrm{SiLU}$ activation from the SwiGLU, which executes on the CGRA while
    the matrix multiplication executes on Gemmini; and (b)~softmax, mapped as two DFGs
    on the CGRA for the reduction and division steps.}
    \label{fig:cgra-mappings}
\end{figure}

% \begin{figure}
%     \centering
%     \includegraphics[width=0.7\linewidth]{images/swiglu_mapping.pdf}
%     \caption{Mapping of the $\mathrm{SiLU}$ on our 4x4 elastic RIKEN CGRA. The $\mathrm{SiLU}$ activation 
%     from the SwiGLU executes on the CGRA and the matrix multiplication executes on Gemmini.}
%     \label{fig:silu-cgra-mapping}
% \end{figure}

\Cref{fig:silu-cgra-mapping} shows the $\mathrm{SiLU}()$ function of \Cref{eq:silu}
mapped onto our 4x4 CGRA. The $g$ in \Cref{eq:silu} corresponds
to a single value of the matrix product $XW_{gate}$, and $u$ is the corresponding single
value of $XW_{up}$ from \Cref{tab:functions-tested}.

The CGRA is surrounded by 32-bit I/O pins, 4 on each side. Only the I/O pins in red, i.e.,
top2, top3, down2 and right0, are active, whereas the gray pins are grounded to zero. Only
5 PEs are active in this mapping, whereas the rest are idle.

The yellow arrows indicate values
that are passed from one PE to another after computation, whereas the blue arrow
shows that the value is passed directly to the FMUL PE at position (3,3).
Constants such as $-1$ are stored in the \texttt{Imm} register of the PE
when the mapping bitstream is loaded into the CGRA using the \texttt{CFG}
instruction from \Cref{tab:swiglu-custom2-instructions}.
\begin{equation}
    y = \frac{g}{1 + \exp(-g)} \cdot u
    \label{eq:silu}
\end{equation}

Notice that the DFG
has two major branches: one branch going through the top2$\to$FMUL$\to$FEXP$\to$FADD$\to$FDIV to
compute $\frac{g}{1+\exp(-g)}$ and the other going through top3$\to$FMUL to compute
the final multiplication with $u$. The left branch
requires $6+3+3+15=27$ cycles, whereas the right one takes only $6$ cycles. While a static
CGRA such as HyCube\cite{Karunaratne2017hycube} or ADRES\cite{Bouwens2008architectural}
would have required inserting extra registers / PEs on the path of the right branch,
elastic CGRAs\cite{Huang2013elastic} such as this one do not have this limitation.

The elastic buffer described in \Cref{sec:matmul-with-cgra} is what drains the back pressure that
the 27-cycle branch exerts on the 6-cycle one.

% \begin{figure}
%     \centering
%     \includegraphics[width=0.7\linewidth]{images/softmax_mapping.pdf}
%     \caption{Softmax mapping on the CGRA. We map two DFGs on the CGRA for softmax: reduction and division.}
%     \label{fig:softmax-mapping}
% \end{figure}

\subsubsection{Mapping the \texorpdfstring{$\mathrm{softmax}$}{softmax}} 
\label{sec:softmax-mapping}

\Cref{fig:softmax-mapping} shows the mapping of the softmax
DFG on our reconfigurable array. This mapping computes both the reduction and element-wise division
loops within the same CGRA context. Here, $m$ is the maximum value of the input data that is
used for normalizing the data to prevent overflow. The maximum computation must iterate over
the entire input data, which will require the projection of a third loop on the CGRA. Therefore,
we choose to compute the maximum value on the Rocket CPU core, and send the value to the
CGRA using the \texttt{SET} instruction shown in \Cref{tab:swiglu-custom2-instructions}.

The reduction loop accumulates a value $acc$ in the internal register of the \texttt{(2,0)}
PE, and then makes that value available to the RoCC controller via the \texttt{left1} I/O pin.
The exact computation performed by this accumulator DFG is shown by \Cref{eq:acc-dfg}.
The $acc$ is read by the Rocket CPU using the \texttt{STAT} instruction. A single accumulation
follows the path FMUL$\to$FADD$\to$FEXP$\to$FADD, and therefore has a latency of 15 clock cycles.
\begin{equation}
    acc = \sum_k \exp(x_k - m)
    \label{eq:acc-dfg}
\end{equation}

Every PE carries an FDIV unit, so the reciprocal could in principle be evaluated on the fabric.
It is not, because $r$ depends on a reduction that has already completed: evaluating it in place
would require a third mapped context, and re-arming the fabric between contexts costs more than
the scalar division it would replace. We therefore treat both $\max()$ and the reciprocal as
mapping choices rather than fabric limitations, and compute them on the Rocket CPU. The CPU
computes $r = 1/acc$ using a standard scalar division operation, and sends the value $r$ back to the \texttt{left2} pin using the
\texttt{SET} instruction. This triggers the computation of the element-wise
multiplication with $r$, given by \Cref{eq:softmax-y-dfg}. A single element-wise exponential and
multiplication follows the path FMUL$\to$FADD$\to$FEXP$\to$FMUL and therefore has a latency of $18$ clock cycles.
\begin{equation}
    y_k = \exp(x_k - m) \cdot r
    \label{eq:softmax-y-dfg}
\end{equation}

% write about the chunking computation mechanism.

For an $N \times N$ input, $\mathrm{softmax}$ normalizes each row independently: every row needs its own maximum and its own sum of exponentials. 
On this design the sum accumulates in a register on the fabric at the PE position \texttt{(2,0)}, and
because that accumulator is FP32, letting it run as a total across rows makes each row's sum a difference of large,
nearly equal numbers --- measured at $1.5 \times 10^{-5}$ relative error. Zeroing it instead, so each partial sum starts from zero, 
reduces the error to $3.8 \times 10^{-7}$ but requires a full fabric reset (the accumulator-read clear bit does not reach the fabric's copy), and a re-arm of the fabric configuration at every chunk boundary.
All results reported in \Cref{sec:results} use the zeroed accumulator. The softmax figures therefore
carry the accuracy of the $3.8\times10^{-7}$ configuration and pay for it in the per-chunk RESET and
ARM of \Cref{tab:softmax-cgra-non-core-analysis}: 4.50 of the 89.3 cycles per element at $N=128$, or
5.0\% of the softmax stage. This is a genuine accuracy-for-performance trade rather than an
incidental overhead --- running the accumulator freely would return those cycles and cost roughly
40x in relative error.

\section{Results}
\label{sec:results}

We report our benchmarks for various functions tested on the architectures shown in \Cref{sec:matmul-with-vector-engine}
and \Cref{sec:matmul-with-cgra}. Cycle-accurate measurements are obtained using Verilator\cite{snyder_verilator}.
DRAMSim2\cite{Rosenfeld2011dramsim2} is run at a clock frequency of 0.66 GHz and the CPU at 0.5 GHz.
To make a fair comparison and keep simulation time manageable, we assume that all the matrices from \Cref{tab:functions-tested} are 
square matrices of size $N \times N$.
To test different problem sizes and scalability, we only vary the $N$ dimension
for each test as $16$, $32$, $64$ and $128$.
The default data type for all computation is IEEE-754 32-bit floating-point numbers (FP32).

Fixing every unit at FP32, the matrix engine included, deserves comment, because production matrix
engines increasingly operate at bf16\cite{kalamkar2019bfloat16} or fp8\cite{micikevicius2022fp8}.
We do so for two reasons. First, the non-linearity itself is normally evaluated in FP32 even when
the matrix multiplication feeding it is not\cite{micikevicius2022fp8}, so FP32 is the precision at
which the vector-unit-versus-fabric comparison is actually made; running the non-linear engines at
a lower precision would measure a different question. Second, an FP32 matrix engine is the slowest
the matrix side can reasonably be relative to the non-linear side. Our Gemmini configuration
delivers 16 FP32 MACs per cycle against four vector lanes, a ratio of four, whereas a low-precision
production engine runs two to three orders of magnitude ahead of its non-linear datapath. The
testbed therefore instantiates a deliberately conservative version of the imbalance that motivates
this work, and every non-linear share we report is a lower bound. That the activation stage already
accounts for 56\% of the SwiGLU block on the ME-VE at $N=128$ under so mild a ratio is what makes
the result interesting: widening the ratio can only increase that share, and with it the value of a
faster non-linear engine. A sweep over the matrix-to-non-linear throughput ratio would quantify how
far the conclusions move and is left as future work. 

Every kernel is executed repeatedly within a single simulation and the first repetition is
discarded as a cold pass; the figures below are taken from the warm repetitions. The
number of warm repetitions differs by campaign because of simulation cost: the SwiGLU sweep
carries nine warm repetitions at $N=16$, 32 and 64 and two at $N=128$, and the attention sweep
carries two at every size. Warm repetitions of the same measurement differ by at most 0.93\%.
Run-to-run variation in an otherwise deterministic RTL simulation comes from cache and DRAM state
carried across repetitions, which is also why the cold pass is excluded.

Two systematic effects should be noted alongside these numbers. The attention drivers exist in an
instrumented \emph{accuracy} build and a lean \emph{timing} build; the timing build measures
0.1--1.1\% fewer cycles than the accuracy build for a computation the two agree on bit-for-bit,
a gap that exceeds the run-to-run spread and whose mechanism we have not identified. All cycle
counts below are quoted from the timing build. Second, the reported instrumentation overhead is
removed by rescaling the per-region shares onto the uninstrumented totals.

\subsection{Performance of SwiGLU on ME-VE vs. ME-RE}
\label{sec:performance-swiglu}

\begin{figure}
    \centering
    \includegraphics[width=0.9\linewidth]{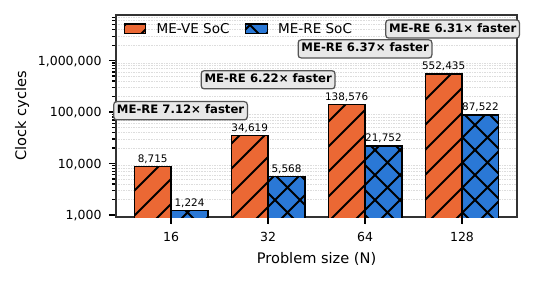}
    \caption{Clock cycles by $SiLU$ on the ME-VE and ME-RE.}
    \label{fig:swiglu-cycles}
\end{figure}

% Table showing total clock cycles and breakdown between matmul and activation.
% Graph showing the activation cycles for swiglu.
\Cref{fig:swiglu-cycles} shows the number of cycles taken for each
problem size and SoC architecture when computing the $\mathrm{SiLU}()$ activation function.
The activation function runs on the RVV vector unit for the ME-VE SoC and on the RIKEN
elastic CGRA for the ME-RE SoC. \Cref{tab:combined-cycles} shows the total cycles
taken by the SwiGLU and the breakdown of matrix multiplication and activation cycles
for each problem size.

As expected, the number of cycles for the matrix
multiplication is nearly identical for the ME-VE and ME-RE SoCs. However, the CGRA shows
significant acceleration of the activation function as shown in \Cref{fig:swiglu-cycles}.
This speedup can be attributed to the fact that the pipelined execution of the CGRA is able to compute the function in fewer cycles per element, in spite of the 4-way parallel execution of the 128-bit SIMD units of the vector processor.

\begin{figure}
    \centering
    \includegraphics[width=0.9\linewidth]{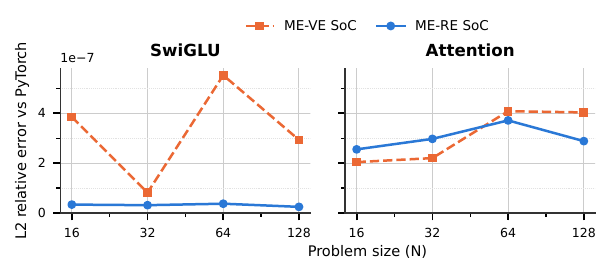}
    \caption{Relative error of SwiGLU (left) and Attention (right) on the ME-VE and 
    ME-RE SoC. The relative error is computed as $\|e - a\|_2/\|a\|_2$ where $e$ is 
    the result vector from the SoC and $a$ is the result vector from a PyTorch
script on the host.}
    \label{fig:combined-relerr}
\end{figure}

The plot on the left of \Cref{fig:combined-relerr} shows the accuracy of the SwiGLU computation
using ME-RE vs. ME-VE SoCs, measured against a PyTorch\cite{Paszke2019pytorch} FP32 reference on
the host. ME-RE stays between $2.5\times10^{-8}$ and $3.7\times10^{-8}$ across all four problem
sizes, whereas ME-VE ranges from $8.2\times10^{-8}$ to $5.5\times10^{-7}$ --- so ME-RE is both
more accurate and more stable in $N$. For reference, the FP32 unit roundoff is
$2^{-24}\approx6\times10^{-8}$, so ME-RE's error is at the level of the arithmetic itself while
ME-VE's exceeds it by up to an order of magnitude. A standalone element-wise comparison of the
CGRA activation against the same reference bounds its error at 1 ULP. The difference is attributable to how the exponential is evaluated on each SoC. The RVV
extension\cite{riscvvector10} used by Saturn has no exponential instruction, so $\exp()$ is
computed by a vector software kernel. That kernel follows the structure used by production
vectorised math libraries\cite{anderson2026correctlyrounded}: the argument is scaled by
$\log_2 e$, split into an integer part $n$ and a fraction $f \in [-\tfrac{1}{2},\tfrac{1}{2}]$
by a float-to-integer-to-float round trip, $2^{f}$ is evaluated by a degree-5 polynomial in Horner
form, and $2^{n}$ is assembled directly into the IEEE-754 exponent field and multiplied back in.
There is no table, no branch and no division; the kernel is 19 vector operations per element, ten
of which are the Horner chain. It is therefore a properly range-reduced implementation rather than
a naive series expansion, and the range reduction is not the source of its error.

It does differ from a production kernel in one respect. The polynomial coefficients are the Taylor
coefficients $(\ln 2)^{i}/i!$ of $2^{f}=e^{f\ln 2}$ rather than a minimax fit over the reduced
interval. At degree 5 this costs accuracy but not time: the Taylor truncation has a worst-case
relative error of $3.2\times10^{-6}$ on $f \in [-\tfrac{1}{2},\tfrac{1}{2}]$, or roughly 27 ULP,
against $7.5\times10^{-8}$ --- under one ULP --- for a minimax polynomial of the same degree and
hence the same instruction count. Since the coefficients are compile-time constants, refitting them
would close most of the accuracy gap at no cycle cost, and we note it as an easy improvement to the
baseline rather than a property of RVV. In contrast, the FloPoCo\cite{Dedinechin2011flopoco} cores
used by our CGRA evaluate the exponential in a single pipelined hardware operator, which is why
ME-RE tracks the reference more closely.

\subsubsection{Performance analysis of SwiGLU}
\label{sec:performance-analysis-swiglu}

\begin{figure}
    \centering
    \includegraphics[width=0.9\linewidth]{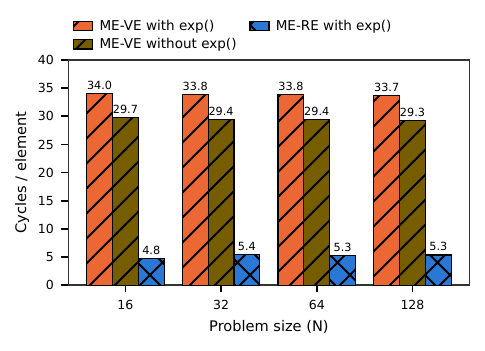}
    \caption{Cycles / element for computing the $SiLU$ for the ME-VE with and without
    the exponential compared to the ME-RE with the exponential. Costs are normalised by the
    $N^2$ elements of the activation tensor, on the same measurements as \Cref{tab:combined-cycles},
    so the ratio between the first and third bars reproduces that table's activation speedup exactly.
    ME-VE is slower than ME-RE even without the exponential calculation, indicating that better pipelining
    leads to faster execution in the ME-RE.}
    \label{fig:swiglu-soc-cycles}
\end{figure}

The lack of exponential instructions in ME-VE also raises the question 
of whether the speedup seen in ME-RE is actually due to better pipelining
or to fewer instructions. \Cref{fig:swiglu-soc-cycles} shows the cycles taken
per element for the ME-VE with and without the $\exp()$ calculation, and for the ME-RE with the
$\exp()$ calculation, when computing the $\mathrm{SiLU}()$. The ME-RE requires fewer cycles per element
than the ME-VE without the exponential. Ablating the exponential from the vector kernel
accounts for a measured 12.7\%, 13.0\%, 13.1\% and 13.2\% of the ME-VE activation stage at
$N=16$, 32, 64 and 128 --- a real but small share, and one that leaves the vector unit between
$5.4\times$ and $6.2\times$ slower per element than the CGRA even when it is removed entirely.
Therefore, the lack of a dedicated exponential
instruction in the RVV instruction set is ruled out as the cause of slower execution
of the vector unit.

Both the CGRA and Saturn are connected to the system bus by a 128-bit TileLink\cite{SiFive2017DiplomaticDP} interface. Even though
the width of the link is the same, Saturn and CGRA differ in how efficient their 
memory instructions are. Saturn issues SIMD load
instructions using \texttt{vle32.v} and SIMD store instructions using \texttt{vse32.v}.
At \texttt{LMUL}$=4$ each such instruction addresses a group of four 128-bit registers and therefore
carries 16 FP32 elements. The DMA of the CGRA, on the other hand,
issues scalar memory instructions that can carry only a single FP32 element each. Saturn's
load / store instructions are therefore 16x more efficient per instruction. Therefore, the efficiency
of memory instructions is heavily in favor of Saturn in spite of the same data width.

% what is the difference between the load instruction issue between the vector and cgra units?

\begin{table*}[t]
\centering
\footnotesize
\setlength{\tabcolsep}{4pt}
\renewcommand{\arraystretch}{1.15}
\caption{Attention and SwiGLU FFN cycle breakdown. The total is read
from an independent counter spanning the whole block rather than summed from the phase counters, so
it does not tally exactly with them. The residual is under 0.02\%.}
\label{tab:combined-cycles}
\begin{tabular}{l|c|cccc|cccc}
\hline
\multirow{3}{*}{\textbf{SoC}} & \multirow{3}{*}{\textbf{$N$}} &
\multicolumn{4}{c|}{\textbf{Attention}} & \multicolumn{4}{c}{\textbf{SwiGLU}} \\
\cline{3-10}
 & & \textbf{Total} & \textbf{Matmul} & \textbf{Activation} & \textbf{Softmax} &
     \textbf{Total} & \textbf{Matmul} & \textbf{Activation} & \textbf{Activation} \\
 & & \textbf{cycles} & \textbf{cycles} & \textbf{cycles} & \textbf{proportion} &
     \textbf{cycles} & \textbf{cycles} & \textbf{cycles} & \textbf{proportion} \\
\hline
\multirow{4}{*}{ME-VE} & 16  &     6,943 &   3,628 &     3,299 & 47.52\% &  10,955 &   2,240 &   8,715 & 79.55\% \\
                        & 32  &    27,610 &  16,582 &    11,022 & 39.92\% &  44,687 &  10,068 &  34,619 & 77.47\% \\
                        & 64  &   143,087 & 103,508 &    39,456 & 27.57\% & 200,890 &  62,314 & 138,576 & 68.98\% \\
                        & 128 &   862,834 & 713,800 &   149,015 & 17.27\% & 983,055 & 430,620 & 552,435 & 56.20\% \\
\hline
\multirow{4}{*}{ME-RE} & 16  &    28,646 &   3,758 &    24,865 & 86.80\% &   3,467 &   2,240 &   1,224 & 35.30\% \\
                        & 32  &   117,022 &  16,838 &   100,091 & 85.53\% &  15,576 &   9,996 &   5,568 & 35.75\% \\
                        & 64  &   461,649 & 103,662 &   357,900 & 77.53\% &  83,892 &  62,147 &  21,752 & 25.93\% \\
                        & 128 & 2,182,546 & 719,890 & 1,462,554 & 67.01\% & 517,529 & 430,004 &  87,522 & 16.91\% \\
\end{tabular}
\end{table*}

% TODO: analysis of the swiglu / CGRA results
% what is the cost per element? 
% how does the pipelining work?
% what if you use SIMD cores in the PEs?
% There are several papers on pipelining analysis of CGRAs. Find out how they
% are useful in this context.
Both engines are pipelined, but they expose parallelism in different ways, and the
comparison turns on which form suits an element-wise loop. The vector engine exploits
\emph{data} parallelism: each instruction presents a group of elements to four FP32
lanes fed by a four-stage pipeline, and the scalar core must issue a fresh instruction
for every such group. The CGRA exploits \emph{pipeline} parallelism: the whole
$\mathrm{SiLU}$ expression is laid out spatially as a 27-cycle datapath which, once
full, retires one element per cycle with no per-element instruction issue at all. For
a loop with no carried dependence and enough elements to fill the datapath, an
initiation interval of one beats issuing groups of elements, because the CGRA pays the
fill cost once per invocation while the vector unit pays an issue cost on every group.
This is why the advantage is largest at the sizes where the fabric stays full and
erodes slightly as $N$ grows and memory traffic begins to dominate. 

% TODO: what is the cause for the pipeline stalls?
% So are there problems in the pipelining? Basically, if you had a single-element
% SIMD engine (vlen=1), would the performance be different? So the main performance
% advantage comes from the parallelism and not from the pipelining? If that is the
% case, will using SIMD inside the CGRA lead to better pipelining + parallel execution?

\subsection{Performance of Attention on ME-VE vs. ME-RE}
\label{sec:performance-attention}

\begin{figure}
    \centering
    \includegraphics[width=0.9\linewidth]{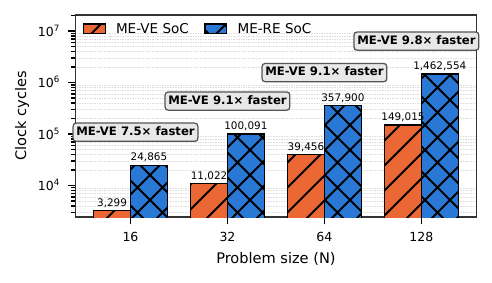}
    \caption{Cycles taken for $softmax$ for ME-VE vs. ME-RE.}
    \label{fig:softmax-soc-cycles}
\end{figure}

\Cref{fig:softmax-soc-cycles} shows the cycles taken for computing the activation function
of the attention computation from \Cref{tab:functions-tested} ($\mathrm{softmax}$) on the ME-VE and
ME-RE. Unlike the SwiGLU results from \Cref{sec:performance-swiglu}, this result shows
that the ME-VE SoC is a much better fit for the $\mathrm{softmax}$ computation across all problem
sizes and by a significant margin, with the ME-VE reaching its maximum speedup of 9.8x at problem size $128$.

The exponential function for the $\mathrm{softmax}$ is calculated differently for the vector processor
and the CGRA, as discussed in \Cref{sec:performance-analysis-swiglu}. This is reflected in the difference
in the relative error between the ME-RE and ME-VE computation as shown in the right plot of
\Cref{fig:combined-relerr}. Neither result should be compared against a fixed threshold: an $L_2$ relative error accumulated
over an $N\times N$ computation grows with $N$, and the PyTorch FP32 reference is itself subject
to the same rounding. We therefore quote both against that reference's own error floor, which is
$2.6\times10^{-7}$ at $N=128$. ME-RE lands at $2.9\times10^{-7}$, or $1.10\times$ the floor, and
ME-VE at $4.0\times10^{-7}$, or $1.54\times$. Both are therefore close to what FP32 arithmetic
permits for this computation, with ME-RE the closer of the two.

\subsubsection{Performance analysis of Attention}
\label{sec:performance-analysis-attention}

\begin{figure}
    \centering
    \includegraphics[width=0.9\linewidth]{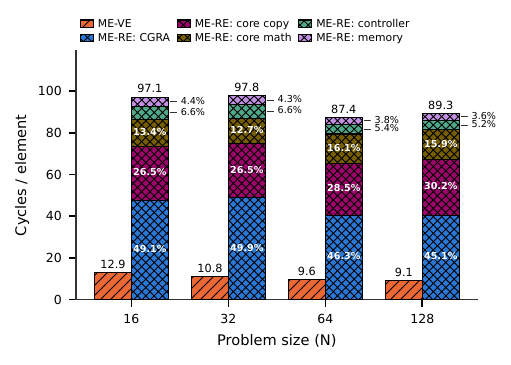}
    \caption{Cycles / element for the activation computation of the Attention calculation
    on the ME-VE compared to the ME-RE. Our mapping gives the ME-VE an advantage over the ME-RE.}
    \label{fig:softmax-cycles-per-element-breakdown}
\end{figure}

The ME-RE is much slower than ME-VE, which warrants further
investigation into the loss of performance.
\Cref{fig:softmax-cycles-per-element-breakdown} shows the cycles / element for computing the $\mathrm{softmax}$
of the $N^2$ matrix from \Cref{tab:functions-tested}. 
The stacked bar on the right of each problem size shows the cycles spent
per element in the ME-RE. Each segment is representative of how many cycles an
element spends doing what kind of work on the SoC. Each element spends cycles going through the CGRA (``ME-RE: CGRA''),
being copied between the CGRA and the Rocket CPU (``ME-RE: core copy''), being computed on the Rocket CPU (``ME-RE: core math''), waiting while the Rocket CPU tells the CGRA what to do (``ME-RE: controller'') or waiting for data reads or writes over the bus from or to the main memory (``ME-RE: memory''). The bar on the left of each problem size shows the cycles / element when the same $\mathrm{softmax}$ computation is done on the vector processor of the ME-VE. The following sections will explore the CGRA
performance in further detail. 

\subsubsection{Math and data copy on the Rocket CPU}
\label{sec:math-data-copy-rocket-cpu}

In \Cref{fig:softmax-cycles-per-element-breakdown}, most of the clock cycles for the ME-RE are spent in the
CGRA (``ME-RE: CGRA'') and the Rocket CPU
(``ME-RE: core math'' and ``ME-RE: core copy''), with both occupying roughly half the clock cycles per element.
We explain in \Cref{sec:softmax-mapping} that the reduction and element-wise
division (multiplication after reciprocal) loops are executed on the CGRA, whereas the $max()$ and the
reciprocal are computed on the Rocket CPU. This incurs significant communication overhead (``ME-RE: core copy'') over
the system bus, which exceeds the time taken for computation.

The communication between the Rocket CPU and the CGRA is serial, with no overlap
of communication and computation. Moreover, each element has to be loaded twice into
the CGRA --- once for computing the reduction loop and a second time for the element-wise
division. This is due to the limited input buffer space in the scratchpad.

These limitations are surmountable with a larger scratch pad and better overlap between
the communication and computation. However, they would only overcome about 30\% of
the cycles (from the ``ME-RE: core copy''). Looking at the time spent inside the CGRA
itself (``ME-RE: CGRA''), the number of cycles of only this part is many times
higher than the total cycles spent in the RVV computation. Improving this metric
can open up significant performance gains for the $\mathrm{softmax}$ on the CGRA.

\subsubsection{Analysis CGRA and non-core computation}

As noted in \Cref{sec:math-data-copy-rocket-cpu}, the Rocket CPU calculation and data movement occupies a significant
number of cycles, but investigating the reasons for the CGRA taking more cycles than
the entire activation computation on the ME-VE can lead to speedups similar to the $\mathrm{SiLU}$
activation in \Cref{sec:performance-swiglu}.

\Cref{tab:softmax-cgra-non-core-analysis} shows where an element spends cycles for the $N=128$ case
from \Cref{fig:softmax-cycles-per-element-breakdown}. The RESET, ARM and RDACC are a part of the
``ME-RE: controller''; FILL and DRAIN are a part of ``ME-RE: memory'', and RUN is entirely ``ME-RE: CGRA''.
The initial configuration of the CGRA is done once per program and is not counted in the total measurement.

The RESET is performed once for every chunk of 24 elements to prevent residual errors caused by stale values.
It is applicable to the reduction loop to ensure that each chunk iteration shows a fresh partial sum,
and for an element-wise kernel to clear out stale values in the pipeline. The RESET causes the 
constants such as $m$ to be erased from the $Imm$ registers, which makes ARM necessary once every chunk.
Specifically for the reduction loop, RDACC is used for clearing and reading the partial sum from
the register in the accumulator PE. The partially accumulated value is stored in a register
of the Rocket CPU, and finally used for computing the reciprocal $r$ which is then used in computing
element-wise loop from \Cref{eq:softmax-y-dfg} on the CGRA.

\begin{table}
\centering
\footnotesize
\setlength{\tabcolsep}{4pt}
\renewcommand{\arraystretch}{1.15}
\caption{Breakdown of the number of cycles an element spends in the CGRA and non-core parts of the SoC.}
\label{tab:softmax-cgra-non-core-analysis}
\begin{tabular}{c|l|c}
\textbf{State} & \multicolumn{1}{c|}{\textbf{Description}} & \begin{tabular}[c]{@{}c@{}}\textbf{Cycles / element}\\\textbf{(N=128)}\end{tabular}  \\ 
\hline
RESET          & Per chunk fabric reset.      & $3.94$   \\
ARM            & \begin{tabular}[c]{@{}l@{}}Re-arm row constants (\$m\$, \$r\$) and~\\stream pointers cleared by the reset.\end{tabular} & $0.56$ \\
RDACC          & \begin{tabular}[c]{@{}l@{}}Read and clear the on-fabric~\\accumulator.\end{tabular}   & $0.14$  \\ 
RUN            & \begin{tabular}[c]{@{}l@{}}Execute one iteration of a chunk of 24~\\elements.\end{tabular} & $40.62$  \\
FILL           & \begin{tabular}[c]{@{}l@{}}Fetch a chunk over DMA and place it~\\in the scratch pad.\end{tabular}  & $2.80$ \\
DRAIN          & Barrier waiting for RUN to finish.  & $<0.01$  \\
\hline
\textbf{Total} & \multicolumn{1}{c}{}  & \textbf{48.06}  \\
\hline
\end{tabular}
\end{table}

Given the enhanced pipeline efficiency of a CGRA compared to a vector processor, the RUN phase should only occupy 
1 clock cycle. However, \Cref{tab:softmax-cgra-non-core-analysis} shows that it requires more than 40x the ideal.
Where does all this time go? Five factors separate that ideal from the 40.62 cycles per
element measured at $N=128$. The first three are multipliers on the number of
tokens a single element costs; the last two are the cycles those tokens then
consume. Table~\ref{tab:fabric-decomp} accumulates them in order. The decomposition
that follows is derived at $N=128$ only, and we do not claim it reproduces at the
other three problem sizes: the prologue factor and the per-invocation costs both
depend on how a row divides into chunks, and at $N \le 24$ a row is a single chunk.
\begin{table}[t]
\centering
\caption{Per-element cost of the softmax fabric at $N=128$ in cycles/element. The
final column is the running total.}
\label{tab:fabric-decomp}
\begin{tabular}{lcr}
\toprule
Step & Factor & Cumulative \\
\midrule
Ideal (II${}=1$)    & ---            & 1.00 \\
Two passes          & $\times 2$     & 2.00 \\
Operand duplication & $\times 2$     & 4.00 \\
Operand prologue    & $\times 1.375$ & 5.50 \\
Streaming rate      & $\times 5.005$ & 27.53 \\
Restart fill        & $+13.02$       & \textbf{40.55} \\
\midrule
Measured            &                & 40.62 \\
\bottomrule
\end{tabular}
\end{table}

\begin{enumerate}
\item \textbf{Two passes ($\times 2$).} Each element crosses the fabric twice:
once to accumulate the sum of exponentials, once for the element wise division.
The reciprocal does not exist until the whole row has been reduced, so the two
passes cannot be fused. This is structural to exact softmax and is the one
factor here that no implementation choice removes.

\item \textbf{Operand duplication ($\times 2$).} The mapped fabric exhibits a
skew between the arms of a fork, so a join pairs operand $i$ on one
arm with operand $i-1$ on the other. Duplicating every operand into two
consecutive tokens makes the skew harmless---a mis-paired token now carries the
same value as its neighbor---at the cost of streaming twice as many tokens.
This is the slack-matching problem familiar from latency-insensitive
design~\cite{Carloni2001,Josipovic2018dynamic,Josipovic2020bufferplacement}, resolved in software rather than by
rebalancing the mapped datapath. Standalone measurement shows the skew cannot
mis-pair operands in the softmax mapping, where two of the three inputs are
row constants presented identically at every token index; the duplication is
inherited from the $\mathrm{SiLU}$ mapping of \Cref{sec:silu_mapping}.

\item \textbf{Operand prologue ($\times 1.375$).} Each run is prefixed with
eight operands equal to the row maximum $m$, absorbing a fault that evaluates
the first operand of every run as $\exp(0)$ irrespective of its input. Each
prologue operand correctly evaluates to $\exp(m-m)=1$, so the prologue
contributes exactly its own length to the accumulated sum---a known constant the
driver subtracts---and the element the fault corrupts is itself a prologue
element whose correct value was $1$ regardless. Eight rather than one because
the DMA engine transfers whole beats and rejects a non-beat-aligned length. A
row of 128 elements is chunked as $[24,24,24,24,24,8]$ and therefore stages 176
operands, the last chunk paying a full prologue for eight real elements.

\item \textbf{Streaming rate ($\times 5.005$).} Each set of duplicated operands occupies the fabric
for 5.005 cycles rather than the one its initiation interval
allows~\cite{Rau1994iterative}, so 22.03 of this step's 27.53 cycles per element is
deficit. We hypothesize that this is a loop-carried
dependence through the on-fabric accumulator, whose serialisation would be
proportional to token count as observed.

\item \textbf{Restart fill ($+13.02$).} The fitted cost is 138.9 cycles per invocation, where one invocation is
one chunk streamed once: a row is chunked as $[24,24,24,24,24,8]$ and each chunk
is streamed once per pass, so a row costs $6 \times 2 = 12$ invocations. The
overhead per row is therefore $12 \times 138.9 = 1667$ cycles, or
$1667/128 = 13.02$ cycles per element.
\end{enumerate}

Of the five, the first three are structural counts, the streaming rate of factor~4 is measured
directly from the fabric's token timestamps, and only the per-invocation cost of factor~5 is
fitted to the measurement. The reconstruction therefore agrees with the measurement to $0.2\%$,
but that agreement is a consistency check on the first four terms rather than an independent
validation of the model as a whole.

% The EB depth is an important parameter that balances the depth of an unbalanced DFG.
% Then write about the scratchpad being there vs. not being there, and the limitation
% of the bandwidth on the number of clock cycles for execution of loops.

\subsubsection{Improving performance for the softmax}

Four properties of the present design account for most of the accelerator's
cost, and removing them would reduce the softmax stage from 89.3 to 58.8 cycles
per element at $N=128$, narrowing the gap against the vector unit from
$9.8\times$ to $6.5\times$. Each corresponds to one of the factors decomposed above.

The dominant one is that the datapath streams every
operand twice (factor 2), a consequence of unequal latency between the arms of a fork in
the mapped graph rather than of the computation; a balanced mapping would
stream each operand once, which also doubles the payload a chunk can hold, taking
the stage to 68.5 cycles per element and accounting for two thirds of the total
reduction. The second is that
the engine detects completion by waiting for its accumulator output to fall
quiet for 96 consecutive cycles, a window that opens only after the last result
has been produced; this quiet window is the largest component of the
per-invocation restart fill of factor 5, and detecting completion from the
expected result count instead yields 64.6. The third is that the first operand
of every run is evaluated incorrectly, which is currently masked by prefixing
each run with eight placeholder operands (factor 3); correcting the fault at its
source yields 62.2. The fourth is
that the scratchpad forces a row to be processed in several chunks, and that although
it is physically split into two halves the present driver does not alternate them,
so each chunk's operands are fetched in sequence with its own execution rather than
during the previous chunk's; the storage for double buffering therefore exists but is
not used as such. Enlarging the scratchpad and alternating the halves reduces the
invocation count over which factor 5 is paid, and yields 58.8.
These interact, and their individual effects must not be summed: each change
that widens a chunk reduces the number of invocations over which the next
change's per-invocation saving is paid.

\section{Related Work}

\subsection{Non-linear operators on spatial accelerators}

PICACHU~\cite{Qin2025picachu} is the closest work --- a CGRA attached to an LLM accelerator for the operators its matrix engine cannot run, and to our knowledge the only published $\mathrm{softmax}$ on a CGRA. Its baseline is not a vector unit sharing the same memory hierarchy, however, whereas we hold the core, L1/L2, DRAM model and matrix engine fixed and vary only the non-linear engine, so the cycle differences in Figures~\ref{fig:swiglu-cycles} and~\ref{fig:softmax-soc-cycles} are attributable to the accelerator alone. A second line lowers the arithmetic cost of non-linearities on an existing datapath: I-BERT~\cite{kim2021bert} and Softermax~\cite{Stevens2022softermax} use integer or reduced-precision approximations, FlashAttention-4~\cite{zadouri2026flashattention4} evaluates the exponential in software on tensor cores, and Gemmini has been extended with dedicated normalization units~\cite{kim2023full}. A third builds fixed-function units for a single non-linearity on reconfigurable substrates: Shatravin et al.~\cite{shatravin2022sigmoid,shatravin2023softmax} for sigmoid and $\mathrm{softmax}$, and Hyft~\cite{xia2024hyft}, a reconfigurable hybrid-format $\mathrm{softmax}$ accelerator. These gain efficiency by specialising to one operator; we instead ask which datapath should be provided at all, and show in Section~\ref{sec:results} that the answer depends on the operator, not on a hardware exponential: the CGRA wins on element-wise $\mathrm{SiLU}$ even with the exponential removed from the vector kernel, yet loses on the reduction-bearing $\mathrm{softmax}$.

\subsection{CGRA architectures and modeling frameworks}

Among open CGRA frameworks --- Morpher~\cite{Wijerathne2022morpher,Juneja2025building}, OpenCGRA~\cite{Tan2020opencgra}, DSAGEN~\cite{Weng2020dsagen}, Pillars~\cite{guo2020pillars}, CCF~\cite{dave2018ccf} and CGRA-ME~\cite{chin2017cgrame,Ragheb2024cgrame2} --- most target integer datapaths or do not carry floating point through mapping and RTL generation; since FP32 $\mathrm{SiLU}$ and $\mathrm{softmax}$ need division and exponentiation, we chose CGRA-ME, whose architecture language lets us instantiate pipelined FloPoCo~\cite{Dedinechin2011flopoco} FP cores and emit synthesizable RTL. Our fabric follows the elastic CGRA model~\cite{Huang2013elastic} as realized in the RIKEN architecture~\cite{Adhi2022exploration}. Statically scheduled CGRAs such as HyCube~\cite{Karunaratne2017hycube} and ADRES~\cite{Bouwens2008architectural} would require balancing the two $\mathrm{SiLU}$ branches (27 vs.\ 6 cycles, Section~\ref{sec:silu_mapping}) with explicit registers at mapping time; elastic handshaking removes that constraint at the cost of back pressure, making buffer placement and sizing~\cite{Lana2021placement,Lines2006slack} a first-order design parameter.

\subsection{Vector units and SoC integration}

We use Saturn~\cite{Zhao2024saturn}, integrated into the Rocket~\cite{asanovic2016rocket} pipeline rather than attached via RoCC like the CGRA and Gemmini~\cite{genc2021gemmini} --- an issue-path asymmetry intrinsic to the two design styles, which is why we equalize everything below the issue point. Decoupled RVV designs such as Ara~\cite{Cavalcante2020ara} target wider datapaths, and some vector architectures offer hardware assistance for the exponential; our ablation in \Cref{sec:performance-analysis-swiglu} shows this would not change the $\mathrm{SiLU}$ outcome, as the vector unit remains several times slower per element than the fabric even without the exponential. Chipyard~\cite{amid2020chipyard} provides the common substrate for our ``same SoC, one variable'' methodology; prior CGRA-versus-SIMD comparisons are analytical or use separate testbeds and nodes, and RISC-V/CGRA integration~\cite{Ling2021} has been shown without a direct comparison against an RVV unit.

\section{Conclusion}

We implemented and tested two SoCs with exactly the same memory hierarchy, matrix multiplication
unit and CPU driver -- one coupled with a RVV\cite{Zhao2024saturn,riscvvector10} 128-bit SIMD processor (``ME-VE``) and 
the other using a elastic CGRA\cite{Adhi2022exploration} (``ME-RE'').
On these SoCs, we tested two widely used feed forward network architectures: SwiGLU and Attention.
The matrix multiplication costs a similar number of clock cycles on both SoCs, but the
cost of the activation computation such as $\mathrm{SiLU}$ and $\mathrm{softmax}$ differs widely.
On the non-linear stage in isolation, the $\mathrm{SiLU}$ reaches a maximum speedup of 7.12x on
the CGRA over RVV, whereas the $\mathrm{softmax}$ reaches a maximum speedup of 9.8x on RVV over
the CGRA. Amortised over the enclosing feed-forward block these maxima are 3.16x in favour of the
CGRA for SwiGLU and 4.24x in favour of RVV for Attention.

The element-wise access of the $\mathrm{SiLU}$ greatly benefits the $\mathrm{SiLU}$ on the CGRA, whereas the multiple
reduction, division and extrema calculation loops play against the CGRA for $\mathrm{softmax}$. We conduct
a thorough investigation into the causes for the slowdown of the $\mathrm{softmax}$ computation on
the CGRA and propose a set of specific improvements that can be made to the CGRA to make it perform faster
than the RVV.

This paper shows that incorporating a CGRA into an AI accelerator shows promise,
but the inherent difficulty of efficiently mapping algorithms on the CGRA can hold it back from
mass adoption. In the future, we plan to develop on our findings and propose efficient architectures
that can leverage the power of CGRAs for a wide variety of computations. We
also plan to perform a more thorough PPA analysis of the ME-VE vs. ME-RE using a standard cell library.

%%%%%%% -- PAPER CONTENT ENDS -- %%%%%%%%

%%%%%%%%% -- BIB STYLE AND FILE -- %%%%%%%%
\bibliographystyle{IEEEtranS}
\bibliography{refs}
%%%%%%%%%%%%%%%%%%%%%%%%%%%%%%%%%%%%

\end{document}